\documentstyle[12pt]{article}
\def\eq{\begin{equation}}
\def\en{\end{equation}}
\newcommand \be  {\begin{equation}}
\newcommand \bea {\begin{eqnarray} \nonumber }
\newcommand \ee  {\end{equation}}
\newcommand \eea {\end{eqnarray}}

\def \bi{\bibitem}

\def\d{{\rm d}}
 \def\(({\left(}
 \def\)){\right)}
\def\bi{\bibitem}

\def \ov{\over}
\def \a{\alpha}
\def \b{\beta}
\def\D{\Delta}
\def \del{\delta}
\def\bQ{{\bf Q}}
\def \d{{\rm d}}
\def \e{{\rm e}}

\def \CN{{\cal N}}
\def \nn{\nonumber}
\def \beqna{\begin{eqnarray}}
\def \eeqna{\end{eqnarray}}
\def \beq{\begin{equation}}
\def \eeq{\end{equation}}
\def \be{\begin{equation}}
\def \ee{\end{equation}}

\def \ov{\over}

\def \ol{\overline}
\def \a{\alpha}

\def \b{\beta}
\def \r{\right}

\def \l{\left}

\def \la{\langle}
\def \ra{\rangle}
\def \Tr{{\rm Tr}}
\def \eps{\epsilon}
\def \ab2{\alpha\beta^2}

\def \la{\langle}
\def \ra{\rangle}

\def \ll{\lambda}
\def \L{\Lambda}

\begin{document}
\begin{center}
{\bf \large On chaos in mean field spin glasses }
\vspace{.5 cm}

{ Silvio Franz(1),
 Muriel Ney-Nifle(2)

\vspace{.5 cm}
 (1) Laboratoire de Physique Th\'eorique,

 Ecole Normale Sup\'erieure\footnote {Unit\'e propre du CNRS,
  associ\'ee
 \`a\ l'Ecole
 Normale Sup\'erieure et \`a\ l'Universit\'e de Paris Sud}

24, Rue Lhomond, 75231 Paris Cedex 05, France
\footnote{ present address:
NORDITA, Blegdamsvej 17,
DK-2100 Copenhagen \O, Denmark;
e-mail: franz@nordita.dk}

\vspace{.5 cm}
 (2) Laboratoire de Physique Th\'eorique et Hautes Energies
\footnote {Laboratoire associ\'e au CNRS},

Batiment 211, Universit\'e de Paris-Sud,

91405 Orsay, France
\footnote{present address : Laboratoire de Physique,
 Ecole Normale Sup\'erieure,
46 all\'ees d'Italie, 69364 Lyon cedex 07, France}
}
\vspace{2 cm}

{\bf Abstract}
\end{center}
We study the correlations between two equilibrium states of
SK  spin glasses
at different temperatures or magnetic fields. The question,
presiously investigated by Kondor and Kondor and V\'egs\"o,
is approached here constraining two copies of the
same system at different external parameters to have a fixed overlap.
We find that imposing an overlap different from the minimal one
implies an extensive cost in free energy.
This confirms
by a different method the Kondor's finding  that equilibrium states
corresponding to different values of the external parameters are completely
uncorrelated.
We also consider the Generalized Random Energy Model of Derrida
as an example of system with strong correlations among states
at different temperatures.
\newpage

\section{Introduction}
The structure of the equilibrium states of  mean field spin glasses
has been widely discussed in the literature \cite{MPV}.
At low temperature ergodicity
is broken, and the contribution to the Boltzmann average comes from
"valleys" separated by infinite barriers. The statistics of the
correlations between the different valleys
for equal values of the external parameters, is one of the
most remarkable  outcomes of the
replica method. In the context of the
Parisi ansatz, it is found that  the  function $q(x)$ is directly
related to the statistics of couples of valleys, while the whole set of states
is organized as an
 ultrametric tree. Much less information is available
about the relations among
the equilibrium states for  different values of the
 external parameters. The object of this paper is an investigation of this
relation.

The states of mean field spin glasses can be
thought as low free energy solution of
TAP equations. Changings in  the external parameters can cause the appearance
or
the disappearance of solutions and modify the relative order of the
 free energy levels.
There are models, for example the $p$-spin spherical spin glass, where
the order of the levels is not affected by changes in temperature
\cite{KPV}. In this case the states at different temperatures are  correlated.
In a model where a change of an external parameter imply a
reshuffling of the states {\it of an extensive amount}, we expect
to find zero correlations between states at different values of
the external parameters.

The problem of the correlations between
low free energy states
at different values of the external parameters has been  addressed for the SK
model by
Kondor \cite{Kondor1} and Kondor and V\'egs\"o \cite{Kondor2}.
Within the frame of the replica method they considered the
partition function of
two realizations of the same system for different external parameters.
In these papers there  were  assumed constant correlations; this
 constant was found
to be zero
in mean field  and  the Gaussian corrections to this situation were computed.

Here we reexamine the question without using this assumption.
Instead  we follow a method introduced in
 \cite{FPV1}, and we take into account in the partition function ($Z$)
 of the two systems
 only these couples of configurations having
 overlap equal to a fixed value
$p_d$.  It was shown in \cite{FPV1}, in the case of two
systems at equal external parameters,
 that if $p_d$ is in the support of the probability function $P(q)$,
the logarithm of the constrained partition function is extensively equal
to that of an unconstrained
system. This result tells simply  that
the  partition function is dominated by these couple of equilibrium
states which
satisfy the constraint.
Conversely, if $p_d$ is out of the support of the $P(q)$ the system
is forced out of equilibrium and this implies
an extensive increase of the free energy  $F=-{\rm log} Z$.

In the present case,
 an extensive increase of  $F$ implies
a reshufflings of the free energy level of an extensive amount, and
zero correlation between states for different external parameters.
The method, relying on the saddle point approximation, is limited to the
computation of the  extensive part of the free energy. So, a zero cost in free
energy density would not  strictly
 imply no reshuffling, but just that the reshuffling is not extensive.
We will refer to "chaos" with respect to an external parameter
to a situation in which the free energy increase is extensive.

The paper is organized as follows:
In section 2 we state the basic definitions and our method.
In section 3 we discuss the problem in the simple case of the Derrida
Generalized Random Energy Model (GREM), where handwaving arguments show
that there is no chaos with temperature in absence of a magnetic field,
and along the lines of constant magnetization. We show how some
modifications in those models produce
chaos with temperature. In section 4 we study the SK model
near the glassy transition. We show that no ultrametric solution exist
for the problems with different magnetic fields or temperatures.
We argue that this is the sign that chaos is present in both cases
and give estimation for the free energy increase.
Finally we draw our conclusions in section 5.

\section{The model}
Let us consider a system composed of two copies
 of a Sherrington Kirkpatrick  (SK) model, having different temperatures and
magnetic fields, $(T_1,h_1)$ and $(T_2,h_2)$ respectively.
The partition function of such a system is
\be
Z=\sum_{ \{S_i^1,S_i^2\} } \exp \l[ \b_1 \sum_{i<j}^{1,N}J_{ij} S_i^1S_j^1 +
h_1\sum_i^{1,N} S_i^1 + \b_2 \sum_{i<j}^{1,N}J_{ij} S_i^2S_j^2 +
h_2\sum_i^{1,N} S_i^2 \r] =Z_1 Z_2.
\label{Z}
\ee
Where the couplings $J_{ij}$ ($i,j=1,...,N$)
are Gaussian independent variables of zero mean and
 variance $1/N$, and the spins $S_i^r$ are Ising variables.
In the following we will improperly call {\it free energy} the quantity
$F=-\log Z$.
To address the question of the correlations between
 the states dominating $Z_1$,
and $Z_2$ respectively, let us
consider  $Z(p_d)$ the sum (\ref{Z})
 restricted to these configurations which verify
\be
p_d={1\ov N}\sum_i S_i^1 S_i^2.
\label{constraint}
\ee
It is clear from the definition that $Z(p_d)\leq Z$.
 It is shown in \cite{FPV1}
in the case $T_1=T_2$ and $h_1=h_2$ that, in the low temperature phase, one has
$(1/N) \log Z=(1/N) \log Z(p_d)$ as soon as $p_d$ is in the support
of the
function $P(q)$ of the free system. This is a consequence of the fact that
the number of valleys dominating the partition function grows less then
exponentially with $N$, as it can be easily understood noting that
$Z=\int \d p_d Z(p_d)$. An increase in free energy at an extensive level,
implies the absence of low-lying states having overlap $p_d$.

In \cite{FPV1} it was shown how to deal with a problem of two coupled
copies with the replica method. We shall not
 repeat here the derivation of
formulae which is completely analogous to that ref.\cite{FPV1},
 but we shall just sketch the results.
Instead of the usual order parameter matrix $Q_{ab}$ of the replica method,
there are 3 matrices $Q^{(1)}_{ab}$, $Q^{(2)}_{ab}$ and $P_{ab}$
representing respectively
\be
Q^{(r)}_{ab}={1\ov N}\sum_{i=1}^N {S_i^r}^a {S_i^r}^b;\;\;\; r=1,2\;\;\;
P_{ab}={1\ov N}\sum_{i=1}^N {S_i^1}^a {S_i^2}^b.
\ee
where $a,b=1,...,n$ and $n$ is "the number of replicas", which as usual has
 to be sent to zero.
The constraint in the partition function implies that the elements $P_{aa}$
have to be  set equal to $p_d$.
Combining $Q^{(1)}$, $Q^{(2)}$, $P$ and its transposed $P^T$ into the matrix
$\bQ=
\l(
\begin{array}{ll}
Q^{(1)} & P\\
P^T & Q^{(2)}
\end{array}
\r)
$, and denoting its elements as $\bQ_{\a\b}$, $\a=(r,a),\;\;\b=(s,b)$
 it is possible to see that for both $T_1$ and $T_2$ near the critical
temperature $T_c=1$ and $h_1$ and $h_2$ small, $\bQ_{\a\b}\sim 1-T_1$ and
 the free energy admits the
expansion up to the fourth order in $T_s-T_c$ ($s=1,2$):
\beqna
F(p_d)&=&-\lim_{N\to\infty} {1\ov N}\log Z(p_d)
\nn
\\
&= &
-\lim_{n\to 0}{1\ov 2n}
\l\{
     \tau_1 \Tr {Q^{(1)}}^2+ \tau_2 \Tr {Q^{(2)}}^2 + 2\tau_{12} \Tr P^2
    +{w\ov 3} \Tr \bQ^3
\r.
\nn
\\
&+&\l.
{u\ov 6} \sum_{\a\b}\bQ_{\a\b}^4
    +{v\ov 4} \Tr \bQ^4 -{y \ov 2} \sum_{\a\b\gamma}\bQ^2_{\a\b}
\bQ^2_{\b\gamma}
\r.
\nn
\\
&+&
\l.
h_1^2 \sum_{ab}Q^{(1)}_{ab}
+ h_2^2 \sum_{ab}Q^{(2)}_{ab}
+ 2h_1 h_2 \sum_{ab}P_{ab}
\r\}
\label{F}
\eeqna
where
\be
\tau_s=(1-T_s^2)/2\;\;\;s=1,2\;\;\;\;
\tau_{12}=(1-T_1 T_2)/2.
\ee
For the "complete" SK model  $y=u=v=w=1$. It is costumary
in the study of the glassy transition to consider a "truncated", (or "reduced")
model in which it is arificially posed $y=v=0$, and retained only the term
$\sum_{\a\b}\bQ_{\a\b}^4$
among all the quartic terms.
 Kondor and V\'egs\"o
have shown recently that this can give rise to instabilities in
considering
couples of systems with different temperatures when the magnetic field
is zero.
We anticipate that the argument showing the presence of
chaos do not depend critically on which of the two models is
used. So, we will use the  complete model to prove the presence of chaos
and we will estimate the free energy increase within the truncated one.
  In any case (\ref{F}) has to be maximized with respect to
the values of the elements of the replica matrices.

The basic object of our investigation will be
the free energy difference
\be
\D F=F(p_d)-F
\ee
where $F$ is the logarithm of the partition function of the two systems
(at two different external parameters)
without constraint, and is equal to
 the sum of the free energies of the two systems.
In the following we will refer to $\D F$ as "free energy excess", and
to "chaos" whenever this quantity is non zero.

Following \cite{FPV1} we will consider here an analytic continuation
to $n\to 0$ for $F(p_d)$ in which each of the matrices $Q^{(r)}$ and $P$
are parametrized according to the Parisi scheme, that is,  specifying
the value of the diagonal elements and a function of the interval [0,1].
\be
Q^{(r)}\to (0,q_r(x));\;\;\;\;
P\to (p_d,p(x))\;\;\; 0\leq x\leq 1.
\ee
The usual restriction of the choice of the Parisi function to the space of
non decreasing function, is here substituted by the requierement of
semi-positive
definiteness of the matrices
\be
\l(
\begin{array}{ll}
q_1'(x) & p'(x)\\
p'(x) & q_2'(x)
\end{array}
\r)
\label{semi}
\ee
for any $x$, where the primes denote derivation with respect to $x$
\cite{FPV2}. In particular, this implies that the functions
$q_s'(x)$ are both positive.

The saddle point equations of maximization of $F_2(p_d)$ of the truncated model
are written in terms of the $q_s(x)$ and $p(x)$ as
\begin{eqnarray}\label{2.1a}
{\delta F\ov \delta q_s(x)}=
2
[\tau_s-\la q_s\ra ]
q_s(x)+
2
[p_d-\la p\ra ]
p(x)-\int_0^x dy
[q_s(x)-q_s(y)]^2\nonumber
\\
 -\int_0^x dy
[p(x)-p(y)]^2
+{2\ov 3} q_s^3(x)+h_s^2=0,
\end{eqnarray}
\begin{eqnarray}\label{2.1b}
{\delta F\ov \delta p(x)}=[\tau_{12}-\la q_1+q_2\ra ]p(x)+
[p_d-\la p\ra ][q_1(x)+q_2(x)]\nonumber
\\
-\int_0^x dy[q_1(x)-q_1(y)+q_2(x)-q_2(y)][p(x)-p(y)]
+{2\ov 3} p^3(x)+h_1h_2=0,
\end{eqnarray}
We also write  for further reference the expression for the derivative
of $F$ with respect to $p_d$.
\begin{equation}\label{2.1c}
{\partial F\ov \partial p_d}=
-
\big[
\tau_{12} p_d-\la p(q_1+q_2)\ra
+{2\ov 3} p_d^3 +h_1h_2
\big]
\end{equation}
Our variational equations differ from the one considered in
reference \cite{Kondor1,Kondor2}
in the fact that there $ p_d$ was taken as a variational
parameter and the function $p(x)$ was constrained to a constant.
We will see in the next section that models without chaos require non
constant $p(x)$.

Before starting the discussion of the maximization of (\ref{F})
and the solutions of (\ref{2.1a},\ref{2.1b}),
we discuss the  correlations between states at different temperature
and magnetic field in the GREM.

\section{The GREM: a model without chaos.}
Let us briefly review the Derrida construction of the Generalized
Random Energy Model. Without any pretension of being exhaustive on this point,
we refer the reader to the original papers on the model \cite{DE,DEGa1,DEGa2}.

In the GREM one considers $2^N$ configurations, associated to the "leaves"
of an ultrametric tree. The tree, is composed of $L$ levels of branching.
At a level $\a$, each branch generates $\CN_\a=\exp({N S_\a})$ new branches, in
such a way that $2^N=\e^{N \sum_\a S_\a}$. To each branch at a level $\a$ is
associated
a random energy so that the total energy
of a configuration is given by $E=\sum_\a \eps_\a$. The $\eps_\a$
are taken as independent Gaussian variables of zero mean and variance
$\ol{\eps_\a^2}=N J_\a^2/2$.
Two configurations are conventionally said to have an overlap
$q_\a$, with $0\le q_\a\le 1$, if they coincide at a level $\a$,
and consequently have the same $\eps_\b$
 for $\b\le \a$. It can be eventually considered a
"continuum limit" of infinite number of levels ($L\rightarrow \infty$)
with infinitesimal spacings, where
$
J_\a\to J(q)\d q\;\;\; S_\a \to S(q)\d q
$
. This exhausts the construction in absence of a
magnetic field. In presence of a magnetic field $h$ one has to associate
magnetization one
to an arbitrarily selected  state, in such a way that the configurations of
magnetization
$m$ are those having an overlap equal to $m$ with this state.
 For these
states the energy gets an extra contribution equal to $-N h m$ where $h$
is the magnetic field.

In the following we will limit our discussion
to the case where $J(q)$ and $S(q)$ increase with $q$, where the
levels associated with
small $q$ freeze at higher temperature than the levels of high $q$.

The absence of chaos with temperature in zero magnetic field is
almost obvious; a change
in temperature do not affect, by  definition of the model, the order of the
energy levels. So, two equilibrium  states at different
 temperatures $T$ and $T'$ can be
strongly correlated. In the same way it is easy to realize that
there is chaos with the magnetic field. Two different magnetic fields, say
$h_1$ and $h_2$, impose to the system to have magnetization respectively
$m_1$ and $m_2$ with $m_1\ne m_2$. By construction the
overlap between two states with such different magnetizations
is equal to $q_{12}=\min \{m_1,m_2\}$. Imposing a different overlap would
bring the magnetizations out of their equilibrium values, implying
an extensive cost in terms of free energy. Furthermore,
in presence of a magnetic field, a
change in temperature implies a change in the magnetization, and again we find
chaos. Thus chaos is clearly  absent along the lines $(T,h)$
of constant magnetization \cite{kur}.

Let us now show
with the aid of the replica method that, for two different temperatures in zero
magnetic field, it is cost-less to impose an overlap $p_d$
in the support of the $P(q)$ function. Note that the
same results could be easily obtained with the Derrida probabilistic technique.
The replicated partition function of the GREM in zero field, in the discrete
formalism is
\be
\ol{Z^n}=
\sum_{ \{ j_{a,s}^\a \}}
\ol{ \exp
         \l( -\sum_{s,a,\a} \b_s \eps_{j_{a,s}^\a}\r)}
\prod_{\a\leq \a_0} \delta_{j_{a,1}^\a,j_{a,2}^\a}
\ee
where the level $\a_0$ corresponds to the overlap $p_d$, $\a =1,\ldots,L$,
$s=1,2$ and $a=1,\ldots,n$ is the indice of replica. Upon
performing the average over the values of the energy levels one gets
\be
\ol{Z^n}=
\sum_{ \{ j_{a,s}^\a \}} \exp \l( {N\ov 4} \sum_{\a,r,s} J_\a^2 \b_{r}\b_s
\sum_{ab}
\delta_{j_{a,r}^\a,j_{b,s}^\a} \r)
\prod_{\a\leq \a_0} \delta_{j_{a,1}^\a,j_{a,2}^\a}.
\ee
To evaluate the partition function we make the following ansatz on the
arrangement of the replicas. We suppose that for the levels
 $\a\leq \a_0$, where $j_{a,1}^\a=j_{a,2}^\a$, the replicas are
divided into $n/x_\a$ groups of $x_\a$ coinciding states (i.e.
 $ j_{a,r}^\a=j_{b,s}^\a
$ for any $r,s$ if $a$ and $b$ are in the same group). For the levels
$\a\geq \a_0$, one has $j_{a,1}^\a\neq j_{a,2}^\a$, and one can divide,
 accordingly
to the same scheme the replicas with $s=1$ and $s=2$ into $n/x_\a^1$ and
$n/x_\a^2$ groups respectively. It is easy to see that the free energy is given
by the expression
\beqna
{1\ov nN}\log \ol{Z^n}&=&
\sum_{\a\leq\a_0}\l[ {S_\a\ov x_\a} +(\b_1+\b_2)^2J_\a^2 x_\a\r]
\nn
\\
&+&
\sum_{\a>\a_0}\l[S_\a ({1\ov x_\a^1}+{1\ov x_\a^2})+J_\a^2(\b_1^2 x_\a^1+
\b_2^2 x_\a^2)\r]
\label{ciccio}
\eeqna
taken at the saddle point aver the various $x$.
Assuming that the level $\a_0$ is frozen at the  temperatures
$T_1$ and $T_2$ one finds upon deriving with respect to $x_\a,
x_\a^1$ and $x_\a^2$
\be
\l\{
\begin{array}{ll}
(\b_1+\b_2)^2 {x_\a^s}^2={S_\a\ov J_\a^2} & \a\leq \a_0 \\
\b_s^2{x_\a^s}^2={S_\a\ov J^2_\a} & \a>
 \a_0\;\;\;{\rm and}\;\; {S_\a / (J_\a^2\b^2_s)} <1\\
x_\a^s=1 & \a>\a_0\;\;\;{\rm and}\;\;{S_\a / (J_\a^2\b^2_s)} >1
\end{array}
\r.
\ee
Substituting in (\ref{ciccio}) one sees easily that
$(1/N) \log \ol{Z^n}= (1/N) [\log \ol{Z_1^n}+\log \ol{Z_2^n}]$,
that is, we find no chaos with temperature.

It is interesting to notice that, in the continuum limit,
inverting the functions $x(q)$ and $x_s(q)$,
 the functions $q_s(x)$ and $p(x)$ take the form:
\be
q_s(x)=
\l\{
\begin{array}{ll}
q_u((\b_1+\b_2)x) & x\leq x_0\\
p_d & x_0\le  x \leq x_s\\
q_u(\b_s x) & x> x_s
\end{array}
\r.
\label{gremq}
\ee
\be
p(x)=
\l\{
\begin{array}{ll}
q_u((\b_1+\b_2)x) & x\leq x_0\\
p_d & x_0\le  x
\end{array}
\r.
\label{gremp}
\ee
where $q_u(\b x)$ is the inverse function of $\b x(q)=S(q)/J^2(q)$, and the
points $x_0$ and $x_s$ are defined by the relations
\be
q_u((\b_1+\b_2)x_0)=q_u(\b_sx_s)=p_d.
\ee
The only solution with a constant $p(x)$ is the one with $p_d=p(x)=0$.

So we have shown that, in a situation where the order of the levels
do not depend on the temperature, imposing an overlap $p_d$ in a
suitable interval do not imply an extensive free energy cost i.e.
there is not chaos with temperature.
The situation is different if the ordering of the state
 depends on the temperature. In the context of the GREM model,
such a dependence can be introduced upon considering one or two of the
following modifications :
\begin{itemize}
{\item choosing temperature dependent $S_\a$ or $J_\a$}
{\item not imposing the  identity of the levels for different
temperatures.}
\end{itemize}
 Here we choose to discuss the simplest possible case, namely the second point.
We take a REM, (i.e. a GREM with only one level, $L=1$) where the Gaussian
 energies depend on $T$, and the correlations are
specified by
\be
\ol{ \eps_j(T_1)\eps_k(T_2)}= \delta_{j,k} {1\ov 2} C(T_1,T_2) N
\ee
where $C(T_1,T_2)\leq J^2=C(T,T)$. Let us compute within the replica formalism,
the partition function of 2 replicas, at temperatures $T_1$ and $T_2$
below the freezing transition,
constrained to be in the same state
\beqna
\ol{Z^n}&=&
\sum_{j^a}\ol{\exp \l[-\b_1 \eps_j^a(T_1)-\b_2 \eps_j^a(T_2)\r]}
\nn\\
&=&
\sum_{j^a} \exp\l\{
N\l[(\b_1^2 +\b_2^2)J^2 +2\b_1\b_2 C(T_1,T_2)\r]
\sum_{ab}\delta_{j^a,j^b} \r\}
\eeqna
Proceeding as above for the GREM (dividing the replicas into groups) one finds
that
\be
{1\ov nN}\log \ol{Z^n}= \sqrt{\log 2 ((\b_1^2+\b_2^2)J^2
 +2\b_1\b_2 C(T_1,T_2))}
\leq
{1\ov nN}\log \ol{Z_1^n} +{1\ov nN}\log \ol{Z_2^n}.
\ee
The equality is recovered for $C(T_1,T_2)=J^2$ which corresponds to
identical levels at the two temperatures.
It would be interesting to understand if this mechanism which produce chaos
with temperature is of any relevance in microscopic models.

\section{Chaos in the SK model.}

Let us now turn to the study of the SK model and
investigate the possibility of absence of chaos.
A possible scenario implying the absence of chaos has been proposed
in \cite{dotsen}. The states at different temperatures are
strongly correlated.  Lowering the temperature the ultrametric  tree
of states undergoes multifuractions in such a way that the states at the  new
temperature are  the descendent  in the tree of the ones at the
old temperature. This is what happens in the GREM, and it seems
reasonable that whenever chaos is absent this must be the correct picture:
the states at different temperatures must be part of the same ultrametric tree.
In this case the total matrix $\bQ_{\a\b}$ should be ultrametric:
for any given three distinct replicas
$\a,\b,\gamma$ one should find
$ \bQ_{\a\b}\geq \min\{\bQ_{\a\gamma},\bQ_{\beta\gamma}\} $.
Specializing the relation to $\a=(1,a)$, $\b=(2,a)$, $\gamma=(1,c)$
, that is, $Q_{1a,2a}=P_{aa}=p_d$ it is easily found that
\be
Q_{1c,2a}=P_{ac}=
\l\{
\begin{array}{ll}
p_d & Q_{1a,1c}\geq p_d
\\
Q_{1a,1c} & Q_{1a,1c}< p_d
\end{array}
\r.
\label{22}
\ee
If we suppose that, as in the case of coinciding external parameters,
the functions $q_s(x)$ and $p(x)$ are continuous \cite{FPV1}, we find
that the condition (\ref{22})  reflects on the functions $q_s(x)$
and $p(x)$ in the following way:
$q_1(x)$ and $q_2(x)$ must be non decreasing in the whole interval $[0,1]$, and
it must exist a point $\ol{x}$ in [0,1] such that
\beqna
 & q_s(x)=p(x)=q_< (x) \;\;\; x\leq \ol{x}
\nn
\\
 & p(x)=p_d \;\;\; x> \ol{x}
\label{um}
\eeqna
By continuity one has $q_< (\ol{x})=p_d$.

The solution (\ref{gremq},\ref{gremp}) for the GREM is obviously of the form
proposed here.
A solution of this form was found in \cite{FPV1} in the case with
  $T_1=T_2=T$ and $h_1=h_2=h$.
It reads
\be
q_1(x)=q_2(x)=
\l\{
\begin{array}{ll}
q_F(2x) & 0\le x\le x_0 /2
\\
p_d & x_0 /2 \le x\le x_0
\\
q_F (x) & x_0\le x\le 1.
\end{array}
\r.
\label{2.8}
\end{equation}
\begin{equation}\label{2.9}
p(x)=
\l\{
\begin{array}{ll}
q_F (2x) & 0\le x\le x_0/2
\\
p_d  & x_0/2\le x\le 1.
\end{array}
\r.
\label{2.9a}
\end{equation}
Where the function $q_F(x)$ is the "free" Parisi function
\begin{equation}\label{2.3}
q_F(x)=
\l\{
\begin{array}{ll}
q_{min}=({3 h^2\ov 4})^{1\ov 3} & 0\le x\le x_{min}
\\
{x\ov 2} & x_{min}\le x\le x_{max}
\\
q_{max}={1-\sqrt{1-4\tau}\ov 2} & x_{max}\le x\le 1
\end{array}
\r.
\end{equation}
where $x_0$ is the point defined by $q_F(x_0)=p_d$ and $x_{min}$ and $x_{max}$
are given by continuity. The interval of
$p_d$ for which this solution is well defined, and that will be considered in
the following, is $q_{min}\leq p_d\leq q_{max}$.
The form  (\ref{2.8},\ref{2.9}) is not limited to this problem; just as a
consequence of ultrametricity
 any model with replica symmetry breaking
admits (\ref{2.8},\ref{2.9}) as solution of the two replicas problem
if the function $q_F(x)$ solves  the single replica one \cite{miatesi}.

The main result of this paper is that as long as $T_1\neq T_2$ or
$h_1\neq h_2$ it does not exist an ultrametric solution of the
kind (\ref{um}) both for the truncated and the complete models of section 2.
A prove of this fact can be given assuming a form of the kind (\ref{um})
and showing that it does not satisfy the saddle point equations.
We postpone this prove to the appendix;
despite its conceptual simplicity the
 prove, already  rather technical for the reduced model,
is complicated by the necessity of using the complete model
if we want a full control of all the terms of
order $\tau_1^4$ in the free energy.

 We conclude that chaos must be present
both in temperature and magnetic field.

 We shall now give some estimate for
the free energy excess to impose the constraint (\ref{constraint})
in various situations.
The solution of equations (\ref{2.1a},\ref{2.1b}) for generic values of
the temperatures, the magnetic fields and $p_d$ is very difficult to find.
We have seen that the situation simplifies for $h_1=h_2$ and $T_1=T_2$
where the solution
for generic $p_d$ is (\ref{2.8},\ref{2.9},\ref{2.3}). Other simple cases,
to be presented below,
are found for $T_1\ne T_2$ and $h_1\ne h_2$ for special values of
$p_d\equiv p_d^0$ which allow for functions $p(x)$ constant with $x$.
It is easy to find that in this last case the system verifies
$\partial F/\partial p_d=0$, that is, the free energy is an extremum
with respect to $p_d$. The only  stable solution is the one
which is a minimum with respect to $p_d$ \cite{Kondor1}, and has a free energy
excess equal to zero\footnote{The reader should not be confused at this point;
we are by no means extremizing $F$ with respect to $p_d$, but
we are claiming that it exists  a special value $p_d^0$ for
which  the free energy has a stable saddle point with $p(x)=constant$.}.
It is easy to find that this solution  must verify:
\be
q_s(x)=q_F(x)\;\;\; {\rm and}\;\;\; p(x)=p_d.
\label{form}
\ee
The values of $p_d$ for which this solution
exists satisfy
\be
\tau_{12} p_d-\la q_1+q_2\ra p_d
+y p_d^3 +h_1h_2=0
\label{ppp}
\ee
which coincides with $\partial F/\partial p_d=0$ (see (\ref{2.1c})).
This solution was first found by Kondor in \cite{Kondor1}.
It is the only solution we found which
has zero free energy excess and it implies
minimal correlations among states
corresponding to different parameters.

We shall
use both the solutions
(\ref{2.8},\ref{2.9})
and (\ref{form},\ref{ppp}) as starting points to compute
 the free energy excess perturbatively in some small parameter.
 We shall consider the three following limit situations:
\begin{itemize}
\item{
      case (1) $T_1=T_2$, $h_1\ne h_2$,
   $p_d=p_d^0+\delta p_d$ and we perturb for small $\delta p_d$
around the solution of
      (\ref{2.1b}) with $p(x)=p_d^0$.
     }
\item{ case (2)
      $T_1=T_2$, $h_2=h_1+\delta h$ fixed $p_d$
and we perturb for small  $\delta h$ around the
solution (\ref{2.8},\ref{2.9}).
}
\item{ case (3) $T_1\neq T_2$, $h_1=h_2$,  $p_d=p_d^0+\del p_d$
and we solve perturbatively in $\del p_d$.}
\end{itemize}
In all cases instead of solving the equations ({\ref{2.1a},\ref{2.1b}}) even in
an approximated form,
we will suitably parameterize the functions $q_s(x)$ and $p(x)$, and
maximize the free energy functional with respect to these parameters.
This variational procedure will enable us to obtain lower bounds for the
free energy excess in the various situations. We expect however to
obtain
the correct order of magnitude of $\Delta F$ as a function of the various
external parameters.
The whole program is analogous to the one pursued in \cite{FPV1}
to compute the free energy excess to have $p_d$ out of the support of
the $P(q)$ for identical parameters or in \cite{FPV2}
to study violations of ultrametricity. We refer the
interested reader to these papers for a presentation more detailed than the
present one.

Let us illustrate as an example the
 case (1). For $p_d=p_d^0+\delta p_d$  ($\delta p_d<<p_d^0$)
we look for functions  $q_s(x)$ and $p(x)$ equal to (\ref{form})
plus some small variations.
These variations, that we call $\del q_s(x)$ and $\del p(x)$,
have to be of order of $\del p_d$ in the saddle point solution.
We choose to parameterize them as follows

\beqna
\label{delq}
&
\delta q_1(x)=
\l\{
\begin{array}{ll}
\del q_1^1 & x < x_m/2\\
\del q_1^2& x_m/2 < x < x_1+\del x_1\\
0 & x>x_1+\del x_1
\end{array}
\r.
\;\;
\delta q_2(x)=
\l\{
\begin{array}{ll}
\del q_2^1 & x < x_m/2\\
\del q_2^2& x_m/2 < x < x_2+\del x_2\\
0 & x>x_2+\del x_2
\end{array}
\r.
&
\nn
\\
&
\label{delp}
\delta p(x)=
\l\{
\begin{array}{ll}
\del p^1 & x < x_m/2\\
\del p^2&  x>x_m/2
\label{delta}
\end{array}
\r.
&
\eeqna
where $x_m/2$ is arbitrarily choosen as the middle of the first
plateau in $q_F$, i.e.
$x_m=\min[x_1,x_2]$ with $ x_s=2 q_{min}^s$ and
$q_{min}^s=(3h_s^2/4)^{1/3}$,  ($s=1,2$).
The various parameters appearing above are
determined by maximization of the free energy functional supposing
self-consistently that they are of order $\delta p_d$. It turns out that
at the lowest order the free energy excess is of order $\delta p_d^2$.
As we are only interested to the lowest order we can minimize the
polynomial of order two obtained expanding up to second order the free
energy functional in all the parameter of order $\delta p_d^2$.
The resulting saddle point equations are linear equations in
the (10) variational parameters, that we solved numerically for
given values of $\tau$, $h_1$ and $h_2$.
In figure
 1 we present the result for the free energy excess at
some values of the external
parameters.
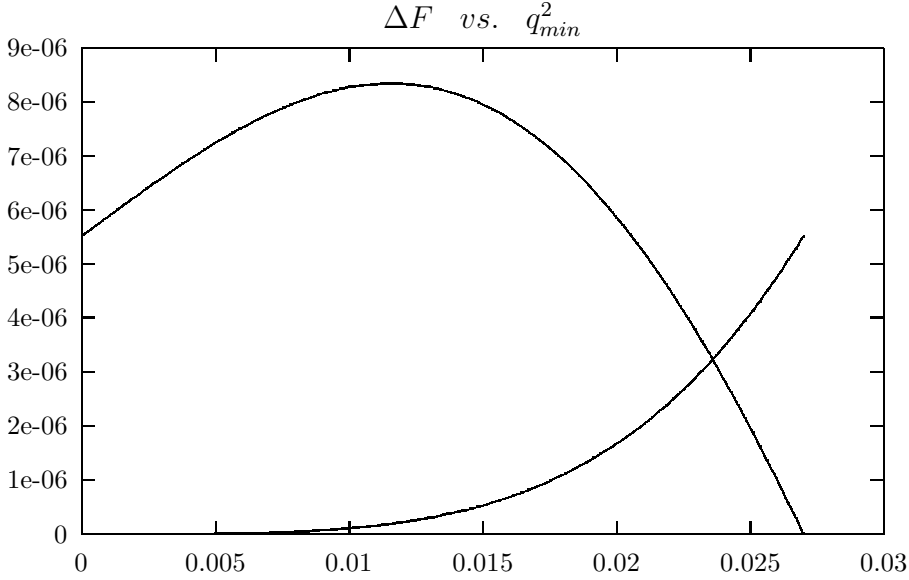
\begin{figure}
\setlength{\unitlength}{0.240900pt}
\ifx\plotpoint\undefined\newsavebox{\plotpoint}\fi
\begin{picture}(1500,900)(0,0)
\font\gnuplot=cmr10 at 10pt
\gnuplot
\sbox{\plotpoint}{\rule[-0.200pt]{0.400pt}{0.400pt}}%
\put(176.0,68.0){\rule[-0.200pt]{303.534pt}{0.400pt}}
\put(176.0,68.0){\rule[-0.200pt]{0.400pt}{184.048pt}}
\put(176.0,68.0){\rule[-0.200pt]{4.818pt}{0.400pt}}
\put(154,68){\makebox(0,0)[r]{0}}
\put(1416.0,68.0){\rule[-0.200pt]{4.818pt}{0.400pt}}
\put(176.0,153.0){\rule[-0.200pt]{4.818pt}{0.400pt}}
\put(154,153){\makebox(0,0)[r]{1e-06}}
\put(1416.0,153.0){\rule[-0.200pt]{4.818pt}{0.400pt}}
\put(176.0,238.0){\rule[-0.200pt]{4.818pt}{0.400pt}}
\put(154,238){\makebox(0,0)[r]{2e-06}}
\put(1416.0,238.0){\rule[-0.200pt]{4.818pt}{0.400pt}}
\put(176.0,323.0){\rule[-0.200pt]{4.818pt}{0.400pt}}
\put(154,323){\makebox(0,0)[r]{3e-06}}
\put(1416.0,323.0){\rule[-0.200pt]{4.818pt}{0.400pt}}
\put(176.0,408.0){\rule[-0.200pt]{4.818pt}{0.400pt}}
\put(154,408){\makebox(0,0)[r]{4e-06}}
\put(1416.0,408.0){\rule[-0.200pt]{4.818pt}{0.400pt}}
\put(176.0,492.0){\rule[-0.200pt]{4.818pt}{0.400pt}}
\put(154,492){\makebox(0,0)[r]{5e-06}}
\put(1416.0,492.0){\rule[-0.200pt]{4.818pt}{0.400pt}}
\put(176.0,577.0){\rule[-0.200pt]{4.818pt}{0.400pt}}
\put(154,577){\makebox(0,0)[r]{6e-06}}
\put(1416.0,577.0){\rule[-0.200pt]{4.818pt}{0.400pt}}
\put(176.0,662.0){\rule[-0.200pt]{4.818pt}{0.400pt}}
\put(154,662){\makebox(0,0)[r]{7e-06}}
\put(1416.0,662.0){\rule[-0.200pt]{4.818pt}{0.400pt}}
\put(176.0,747.0){\rule[-0.200pt]{4.818pt}{0.400pt}}
\put(154,747){\makebox(0,0)[r]{8e-06}}
\put(1416.0,747.0){\rule[-0.200pt]{4.818pt}{0.400pt}}
\put(176.0,832.0){\rule[-0.200pt]{4.818pt}{0.400pt}}
\put(154,832){\makebox(0,0)[r]{9e-06}}
\put(1416.0,832.0){\rule[-0.200pt]{4.818pt}{0.400pt}}
\put(176.0,68.0){\rule[-0.200pt]{0.400pt}{4.818pt}}
\put(176,23){\makebox(0,0){0}}
\put(176.0,812.0){\rule[-0.200pt]{0.400pt}{4.818pt}}
\put(386.0,68.0){\rule[-0.200pt]{0.400pt}{4.818pt}}
\put(386,23){\makebox(0,0){0.005}}
\put(386.0,812.0){\rule[-0.200pt]{0.400pt}{4.818pt}}
\put(596.0,68.0){\rule[-0.200pt]{0.400pt}{4.818pt}}
\put(596,23){\makebox(0,0){0.01}}
\put(596.0,812.0){\rule[-0.200pt]{0.400pt}{4.818pt}}
\put(806.0,68.0){\rule[-0.200pt]{0.400pt}{4.818pt}}
\put(806,23){\makebox(0,0){0.015}}
\put(806.0,812.0){\rule[-0.200pt]{0.400pt}{4.818pt}}
\put(1016.0,68.0){\rule[-0.200pt]{0.400pt}{4.818pt}}
\put(1016,23){\makebox(0,0){0.02}}
\put(1016.0,812.0){\rule[-0.200pt]{0.400pt}{4.818pt}}
\put(1226.0,68.0){\rule[-0.200pt]{0.400pt}{4.818pt}}
\put(1226,23){\makebox(0,0){0.025}}
\put(1226.0,812.0){\rule[-0.200pt]{0.400pt}{4.818pt}}
\put(1436.0,68.0){\rule[-0.200pt]{0.400pt}{4.818pt}}
\put(1436,23){\makebox(0,0){0.03}}
\put(1436.0,812.0){\rule[-0.200pt]{0.400pt}{4.818pt}}
\put(176.0,68.0){\rule[-0.200pt]{303.534pt}{0.400pt}}
\put(1436.0,68.0){\rule[-0.200pt]{0.400pt}{184.048pt}}
\put(176.0,832.0){\rule[-0.200pt]{303.534pt}{0.400pt}}
\put(806,877){\makebox(0,0){$\Delta F\;\;\; vs.\;\;\; q_{min}^2$}}
\put(176.0,68.0){\rule[-0.200pt]{0.400pt}{184.048pt}}
\put(176,68){\usebox{\plotpoint}}
\put(380,67.67){\rule{2.650pt}{0.400pt}}
\multiput(380.00,67.17)(5.500,1.000){2}{\rule{1.325pt}{0.400pt}}
\put(176.0,68.0){\rule[-0.200pt]{49.144pt}{0.400pt}}
\put(437,68.67){\rule{2.650pt}{0.400pt}}
\multiput(437.00,68.17)(5.500,1.000){2}{\rule{1.325pt}{0.400pt}}
\put(391.0,69.0){\rule[-0.200pt]{11.081pt}{0.400pt}}
\put(471,69.67){\rule{2.650pt}{0.400pt}}
\multiput(471.00,69.17)(5.500,1.000){2}{\rule{1.325pt}{0.400pt}}
\put(448.0,70.0){\rule[-0.200pt]{5.541pt}{0.400pt}}
\put(505,70.67){\rule{2.650pt}{0.400pt}}
\multiput(505.00,70.17)(5.500,1.000){2}{\rule{1.325pt}{0.400pt}}
\put(482.0,71.0){\rule[-0.200pt]{5.541pt}{0.400pt}}
\put(528,71.67){\rule{2.650pt}{0.400pt}}
\multiput(528.00,71.17)(5.500,1.000){2}{\rule{1.325pt}{0.400pt}}
\put(539,72.67){\rule{2.650pt}{0.400pt}}
\multiput(539.00,72.17)(5.500,1.000){2}{\rule{1.325pt}{0.400pt}}
\put(516.0,72.0){\rule[-0.200pt]{2.891pt}{0.400pt}}
\put(562,73.67){\rule{2.650pt}{0.400pt}}
\multiput(562.00,73.17)(5.500,1.000){2}{\rule{1.325pt}{0.400pt}}
\put(573,74.67){\rule{2.650pt}{0.400pt}}
\multiput(573.00,74.17)(5.500,1.000){2}{\rule{1.325pt}{0.400pt}}
\put(584,75.67){\rule{2.891pt}{0.400pt}}
\multiput(584.00,75.17)(6.000,1.000){2}{\rule{1.445pt}{0.400pt}}
\put(596,76.67){\rule{2.650pt}{0.400pt}}
\multiput(596.00,76.17)(5.500,1.000){2}{\rule{1.325pt}{0.400pt}}
\put(607,77.67){\rule{2.650pt}{0.400pt}}
\multiput(607.00,77.17)(5.500,1.000){2}{\rule{1.325pt}{0.400pt}}
\put(618,78.67){\rule{2.891pt}{0.400pt}}
\multiput(618.00,78.17)(6.000,1.000){2}{\rule{1.445pt}{0.400pt}}
\put(630,79.67){\rule{2.650pt}{0.400pt}}
\multiput(630.00,79.17)(5.500,1.000){2}{\rule{1.325pt}{0.400pt}}
\put(641,81.17){\rule{2.300pt}{0.400pt}}
\multiput(641.00,80.17)(6.226,2.000){2}{\rule{1.150pt}{0.400pt}}
\put(652,82.67){\rule{2.891pt}{0.400pt}}
\multiput(652.00,82.17)(6.000,1.000){2}{\rule{1.445pt}{0.400pt}}
\put(664,84.17){\rule{2.300pt}{0.400pt}}
\multiput(664.00,83.17)(6.226,2.000){2}{\rule{1.150pt}{0.400pt}}
\put(675,85.67){\rule{2.650pt}{0.400pt}}
\multiput(675.00,85.17)(5.500,1.000){2}{\rule{1.325pt}{0.400pt}}
\put(686,87.17){\rule{2.500pt}{0.400pt}}
\multiput(686.00,86.17)(6.811,2.000){2}{\rule{1.250pt}{0.400pt}}
\put(698,89.17){\rule{2.300pt}{0.400pt}}
\multiput(698.00,88.17)(6.226,2.000){2}{\rule{1.150pt}{0.400pt}}
\put(709,91.17){\rule{2.300pt}{0.400pt}}
\multiput(709.00,90.17)(6.226,2.000){2}{\rule{1.150pt}{0.400pt}}
\put(720,93.17){\rule{2.500pt}{0.400pt}}
\multiput(720.00,92.17)(6.811,2.000){2}{\rule{1.250pt}{0.400pt}}
\multiput(732.00,95.61)(2.248,0.447){3}{\rule{1.567pt}{0.108pt}}
\multiput(732.00,94.17)(7.748,3.000){2}{\rule{0.783pt}{0.400pt}}
\put(743,98.17){\rule{2.300pt}{0.400pt}}
\multiput(743.00,97.17)(6.226,2.000){2}{\rule{1.150pt}{0.400pt}}
\multiput(754.00,100.61)(2.472,0.447){3}{\rule{1.700pt}{0.108pt}}
\multiput(754.00,99.17)(8.472,3.000){2}{\rule{0.850pt}{0.400pt}}
\put(766,103.17){\rule{2.300pt}{0.400pt}}
\multiput(766.00,102.17)(6.226,2.000){2}{\rule{1.150pt}{0.400pt}}
\multiput(777.00,105.61)(2.248,0.447){3}{\rule{1.567pt}{0.108pt}}
\multiput(777.00,104.17)(7.748,3.000){2}{\rule{0.783pt}{0.400pt}}
\multiput(788.00,108.61)(2.472,0.447){3}{\rule{1.700pt}{0.108pt}}
\multiput(788.00,107.17)(8.472,3.000){2}{\rule{0.850pt}{0.400pt}}
\multiput(800.00,111.61)(2.248,0.447){3}{\rule{1.567pt}{0.108pt}}
\multiput(800.00,110.17)(7.748,3.000){2}{\rule{0.783pt}{0.400pt}}
\multiput(811.00,114.60)(1.505,0.468){5}{\rule{1.200pt}{0.113pt}}
\multiput(811.00,113.17)(8.509,4.000){2}{\rule{0.600pt}{0.400pt}}
\multiput(822.00,118.61)(2.472,0.447){3}{\rule{1.700pt}{0.108pt}}
\multiput(822.00,117.17)(8.472,3.000){2}{\rule{0.850pt}{0.400pt}}
\multiput(834.00,121.60)(1.505,0.468){5}{\rule{1.200pt}{0.113pt}}
\multiput(834.00,120.17)(8.509,4.000){2}{\rule{0.600pt}{0.400pt}}
\multiput(845.00,125.60)(1.505,0.468){5}{\rule{1.200pt}{0.113pt}}
\multiput(845.00,124.17)(8.509,4.000){2}{\rule{0.600pt}{0.400pt}}
\multiput(856.00,129.60)(1.651,0.468){5}{\rule{1.300pt}{0.113pt}}
\multiput(856.00,128.17)(9.302,4.000){2}{\rule{0.650pt}{0.400pt}}
\multiput(868.00,133.59)(1.155,0.477){7}{\rule{0.980pt}{0.115pt}}
\multiput(868.00,132.17)(8.966,5.000){2}{\rule{0.490pt}{0.400pt}}
\multiput(879.00,138.60)(1.505,0.468){5}{\rule{1.200pt}{0.113pt}}
\multiput(879.00,137.17)(8.509,4.000){2}{\rule{0.600pt}{0.400pt}}
\multiput(890.00,142.59)(1.267,0.477){7}{\rule{1.060pt}{0.115pt}}
\multiput(890.00,141.17)(9.800,5.000){2}{\rule{0.530pt}{0.400pt}}
\multiput(902.00,147.59)(1.155,0.477){7}{\rule{0.980pt}{0.115pt}}
\multiput(902.00,146.17)(8.966,5.000){2}{\rule{0.490pt}{0.400pt}}
\multiput(913.00,152.59)(0.943,0.482){9}{\rule{0.833pt}{0.116pt}}
\multiput(913.00,151.17)(9.270,6.000){2}{\rule{0.417pt}{0.400pt}}
\multiput(924.00,158.59)(1.267,0.477){7}{\rule{1.060pt}{0.115pt}}
\multiput(924.00,157.17)(9.800,5.000){2}{\rule{0.530pt}{0.400pt}}
\multiput(936.00,163.59)(0.943,0.482){9}{\rule{0.833pt}{0.116pt}}
\multiput(936.00,162.17)(9.270,6.000){2}{\rule{0.417pt}{0.400pt}}
\multiput(947.00,169.59)(0.943,0.482){9}{\rule{0.833pt}{0.116pt}}
\multiput(947.00,168.17)(9.270,6.000){2}{\rule{0.417pt}{0.400pt}}
\multiput(958.00,175.59)(1.033,0.482){9}{\rule{0.900pt}{0.116pt}}
\multiput(958.00,174.17)(10.132,6.000){2}{\rule{0.450pt}{0.400pt}}
\multiput(970.00,181.59)(0.798,0.485){11}{\rule{0.729pt}{0.117pt}}
\multiput(970.00,180.17)(9.488,7.000){2}{\rule{0.364pt}{0.400pt}}
\multiput(981.00,188.59)(0.798,0.485){11}{\rule{0.729pt}{0.117pt}}
\multiput(981.00,187.17)(9.488,7.000){2}{\rule{0.364pt}{0.400pt}}
\multiput(992.00,195.59)(0.874,0.485){11}{\rule{0.786pt}{0.117pt}}
\multiput(992.00,194.17)(10.369,7.000){2}{\rule{0.393pt}{0.400pt}}
\multiput(1004.00,202.59)(0.798,0.485){11}{\rule{0.729pt}{0.117pt}}
\multiput(1004.00,201.17)(9.488,7.000){2}{\rule{0.364pt}{0.400pt}}
\multiput(1015.00,209.59)(0.758,0.488){13}{\rule{0.700pt}{0.117pt}}
\multiput(1015.00,208.17)(10.547,8.000){2}{\rule{0.350pt}{0.400pt}}
\multiput(1027.00,217.59)(0.692,0.488){13}{\rule{0.650pt}{0.117pt}}
\multiput(1027.00,216.17)(9.651,8.000){2}{\rule{0.325pt}{0.400pt}}
\multiput(1038.00,225.59)(0.692,0.488){13}{\rule{0.650pt}{0.117pt}}
\multiput(1038.00,224.17)(9.651,8.000){2}{\rule{0.325pt}{0.400pt}}
\multiput(1049.00,233.59)(0.669,0.489){15}{\rule{0.633pt}{0.118pt}}
\multiput(1049.00,232.17)(10.685,9.000){2}{\rule{0.317pt}{0.400pt}}
\multiput(1061.00,242.59)(0.611,0.489){15}{\rule{0.589pt}{0.118pt}}
\multiput(1061.00,241.17)(9.778,9.000){2}{\rule{0.294pt}{0.400pt}}
\multiput(1072.00,251.58)(0.547,0.491){17}{\rule{0.540pt}{0.118pt}}
\multiput(1072.00,250.17)(9.879,10.000){2}{\rule{0.270pt}{0.400pt}}
\multiput(1083.00,261.59)(0.669,0.489){15}{\rule{0.633pt}{0.118pt}}
\multiput(1083.00,260.17)(10.685,9.000){2}{\rule{0.317pt}{0.400pt}}
\multiput(1095.00,270.58)(0.496,0.492){19}{\rule{0.500pt}{0.118pt}}
\multiput(1095.00,269.17)(9.962,11.000){2}{\rule{0.250pt}{0.400pt}}
\multiput(1106.00,281.58)(0.547,0.491){17}{\rule{0.540pt}{0.118pt}}
\multiput(1106.00,280.17)(9.879,10.000){2}{\rule{0.270pt}{0.400pt}}
\multiput(1117.00,291.58)(0.543,0.492){19}{\rule{0.536pt}{0.118pt}}
\multiput(1117.00,290.17)(10.887,11.000){2}{\rule{0.268pt}{0.400pt}}
\multiput(1129.00,302.58)(0.496,0.492){19}{\rule{0.500pt}{0.118pt}}
\multiput(1129.00,301.17)(9.962,11.000){2}{\rule{0.250pt}{0.400pt}}
\multiput(1140.58,313.00)(0.492,0.543){19}{\rule{0.118pt}{0.536pt}}
\multiput(1139.17,313.00)(11.000,10.887){2}{\rule{0.400pt}{0.268pt}}
\multiput(1151.00,325.58)(0.496,0.492){21}{\rule{0.500pt}{0.119pt}}
\multiput(1151.00,324.17)(10.962,12.000){2}{\rule{0.250pt}{0.400pt}}
\multiput(1163.58,337.00)(0.492,0.543){19}{\rule{0.118pt}{0.536pt}}
\multiput(1162.17,337.00)(11.000,10.887){2}{\rule{0.400pt}{0.268pt}}
\multiput(1174.58,349.00)(0.492,0.590){19}{\rule{0.118pt}{0.573pt}}
\multiput(1173.17,349.00)(11.000,11.811){2}{\rule{0.400pt}{0.286pt}}
\multiput(1185.58,362.00)(0.492,0.582){21}{\rule{0.119pt}{0.567pt}}
\multiput(1184.17,362.00)(12.000,12.824){2}{\rule{0.400pt}{0.283pt}}
\multiput(1197.58,376.00)(0.492,0.637){19}{\rule{0.118pt}{0.609pt}}
\multiput(1196.17,376.00)(11.000,12.736){2}{\rule{0.400pt}{0.305pt}}
\multiput(1208.58,390.00)(0.492,0.637){19}{\rule{0.118pt}{0.609pt}}
\multiput(1207.17,390.00)(11.000,12.736){2}{\rule{0.400pt}{0.305pt}}
\multiput(1219.58,404.00)(0.492,0.625){21}{\rule{0.119pt}{0.600pt}}
\multiput(1218.17,404.00)(12.000,13.755){2}{\rule{0.400pt}{0.300pt}}
\multiput(1231.58,419.00)(0.492,0.684){19}{\rule{0.118pt}{0.645pt}}
\multiput(1230.17,419.00)(11.000,13.660){2}{\rule{0.400pt}{0.323pt}}
\multiput(1242.58,434.00)(0.492,0.732){19}{\rule{0.118pt}{0.682pt}}
\multiput(1241.17,434.00)(11.000,14.585){2}{\rule{0.400pt}{0.341pt}}
\multiput(1253.58,450.00)(0.492,0.669){21}{\rule{0.119pt}{0.633pt}}
\multiput(1252.17,450.00)(12.000,14.685){2}{\rule{0.400pt}{0.317pt}}
\multiput(1265.58,466.00)(0.492,0.732){19}{\rule{0.118pt}{0.682pt}}
\multiput(1264.17,466.00)(11.000,14.585){2}{\rule{0.400pt}{0.341pt}}
\multiput(1276.58,482.00)(0.492,0.826){19}{\rule{0.118pt}{0.755pt}}
\multiput(1275.17,482.00)(11.000,16.434){2}{\rule{0.400pt}{0.377pt}}
\multiput(1287.58,500.00)(0.492,0.712){21}{\rule{0.119pt}{0.667pt}}
\multiput(1286.17,500.00)(12.000,15.616){2}{\rule{0.400pt}{0.333pt}}
\multiput(1299.58,517.00)(0.492,0.873){19}{\rule{0.118pt}{0.791pt}}
\multiput(1298.17,517.00)(11.000,17.358){2}{\rule{0.400pt}{0.395pt}}
\put(550.0,74.0){\rule[-0.200pt]{2.891pt}{0.400pt}}
\put(176,536){\usebox{\plotpoint}}
\multiput(176.00,536.59)(0.692,0.488){13}{\rule{0.650pt}{0.117pt}}
\multiput(176.00,535.17)(9.651,8.000){2}{\rule{0.325pt}{0.400pt}}
\multiput(187.00,544.59)(0.669,0.489){15}{\rule{0.633pt}{0.118pt}}
\multiput(187.00,543.17)(10.685,9.000){2}{\rule{0.317pt}{0.400pt}}
\multiput(199.00,553.59)(0.692,0.488){13}{\rule{0.650pt}{0.117pt}}
\multiput(199.00,552.17)(9.651,8.000){2}{\rule{0.325pt}{0.400pt}}
\multiput(210.00,561.59)(0.692,0.488){13}{\rule{0.650pt}{0.117pt}}
\multiput(210.00,560.17)(9.651,8.000){2}{\rule{0.325pt}{0.400pt}}
\multiput(221.00,569.59)(0.669,0.489){15}{\rule{0.633pt}{0.118pt}}
\multiput(221.00,568.17)(10.685,9.000){2}{\rule{0.317pt}{0.400pt}}
\multiput(233.00,578.59)(0.692,0.488){13}{\rule{0.650pt}{0.117pt}}
\multiput(233.00,577.17)(9.651,8.000){2}{\rule{0.325pt}{0.400pt}}
\multiput(244.00,586.59)(0.692,0.488){13}{\rule{0.650pt}{0.117pt}}
\multiput(244.00,585.17)(9.651,8.000){2}{\rule{0.325pt}{0.400pt}}
\multiput(255.00,594.59)(0.758,0.488){13}{\rule{0.700pt}{0.117pt}}
\multiput(255.00,593.17)(10.547,8.000){2}{\rule{0.350pt}{0.400pt}}
\multiput(267.00,602.59)(0.611,0.489){15}{\rule{0.589pt}{0.118pt}}
\multiput(267.00,601.17)(9.778,9.000){2}{\rule{0.294pt}{0.400pt}}
\multiput(278.00,611.59)(0.692,0.488){13}{\rule{0.650pt}{0.117pt}}
\multiput(278.00,610.17)(9.651,8.000){2}{\rule{0.325pt}{0.400pt}}
\multiput(289.00,619.59)(0.874,0.485){11}{\rule{0.786pt}{0.117pt}}
\multiput(289.00,618.17)(10.369,7.000){2}{\rule{0.393pt}{0.400pt}}
\multiput(301.00,626.59)(0.692,0.488){13}{\rule{0.650pt}{0.117pt}}
\multiput(301.00,625.17)(9.651,8.000){2}{\rule{0.325pt}{0.400pt}}
\multiput(312.00,634.59)(0.692,0.488){13}{\rule{0.650pt}{0.117pt}}
\multiput(312.00,633.17)(9.651,8.000){2}{\rule{0.325pt}{0.400pt}}
\multiput(323.00,642.59)(0.758,0.488){13}{\rule{0.700pt}{0.117pt}}
\multiput(323.00,641.17)(10.547,8.000){2}{\rule{0.350pt}{0.400pt}}
\multiput(335.00,650.59)(0.798,0.485){11}{\rule{0.729pt}{0.117pt}}
\multiput(335.00,649.17)(9.488,7.000){2}{\rule{0.364pt}{0.400pt}}
\multiput(346.00,657.59)(0.798,0.485){11}{\rule{0.729pt}{0.117pt}}
\multiput(346.00,656.17)(9.488,7.000){2}{\rule{0.364pt}{0.400pt}}
\multiput(357.00,664.59)(0.758,0.488){13}{\rule{0.700pt}{0.117pt}}
\multiput(357.00,663.17)(10.547,8.000){2}{\rule{0.350pt}{0.400pt}}
\multiput(369.00,672.59)(0.798,0.485){11}{\rule{0.729pt}{0.117pt}}
\multiput(369.00,671.17)(9.488,7.000){2}{\rule{0.364pt}{0.400pt}}
\multiput(380.00,679.59)(0.798,0.485){11}{\rule{0.729pt}{0.117pt}}
\multiput(380.00,678.17)(9.488,7.000){2}{\rule{0.364pt}{0.400pt}}
\multiput(391.00,686.59)(1.033,0.482){9}{\rule{0.900pt}{0.116pt}}
\multiput(391.00,685.17)(10.132,6.000){2}{\rule{0.450pt}{0.400pt}}
\multiput(403.00,692.59)(0.798,0.485){11}{\rule{0.729pt}{0.117pt}}
\multiput(403.00,691.17)(9.488,7.000){2}{\rule{0.364pt}{0.400pt}}
\multiput(414.00,699.59)(0.943,0.482){9}{\rule{0.833pt}{0.116pt}}
\multiput(414.00,698.17)(9.270,6.000){2}{\rule{0.417pt}{0.400pt}}
\multiput(425.00,705.59)(1.033,0.482){9}{\rule{0.900pt}{0.116pt}}
\multiput(425.00,704.17)(10.132,6.000){2}{\rule{0.450pt}{0.400pt}}
\multiput(437.00,711.59)(0.943,0.482){9}{\rule{0.833pt}{0.116pt}}
\multiput(437.00,710.17)(9.270,6.000){2}{\rule{0.417pt}{0.400pt}}
\multiput(448.00,717.59)(1.033,0.482){9}{\rule{0.900pt}{0.116pt}}
\multiput(448.00,716.17)(10.132,6.000){2}{\rule{0.450pt}{0.400pt}}
\multiput(460.00,723.59)(1.155,0.477){7}{\rule{0.980pt}{0.115pt}}
\multiput(460.00,722.17)(8.966,5.000){2}{\rule{0.490pt}{0.400pt}}
\multiput(471.00,728.59)(1.155,0.477){7}{\rule{0.980pt}{0.115pt}}
\multiput(471.00,727.17)(8.966,5.000){2}{\rule{0.490pt}{0.400pt}}
\multiput(482.00,733.59)(1.267,0.477){7}{\rule{1.060pt}{0.115pt}}
\multiput(482.00,732.17)(9.800,5.000){2}{\rule{0.530pt}{0.400pt}}
\multiput(494.00,738.59)(1.155,0.477){7}{\rule{0.980pt}{0.115pt}}
\multiput(494.00,737.17)(8.966,5.000){2}{\rule{0.490pt}{0.400pt}}
\multiput(505.00,743.60)(1.505,0.468){5}{\rule{1.200pt}{0.113pt}}
\multiput(505.00,742.17)(8.509,4.000){2}{\rule{0.600pt}{0.400pt}}
\multiput(516.00,747.60)(1.651,0.468){5}{\rule{1.300pt}{0.113pt}}
\multiput(516.00,746.17)(9.302,4.000){2}{\rule{0.650pt}{0.400pt}}
\multiput(528.00,751.60)(1.505,0.468){5}{\rule{1.200pt}{0.113pt}}
\multiput(528.00,750.17)(8.509,4.000){2}{\rule{0.600pt}{0.400pt}}
\multiput(539.00,755.60)(1.505,0.468){5}{\rule{1.200pt}{0.113pt}}
\multiput(539.00,754.17)(8.509,4.000){2}{\rule{0.600pt}{0.400pt}}
\multiput(550.00,759.61)(2.472,0.447){3}{\rule{1.700pt}{0.108pt}}
\multiput(550.00,758.17)(8.472,3.000){2}{\rule{0.850pt}{0.400pt}}
\multiput(562.00,762.61)(2.248,0.447){3}{\rule{1.567pt}{0.108pt}}
\multiput(562.00,761.17)(7.748,3.000){2}{\rule{0.783pt}{0.400pt}}
\multiput(573.00,765.61)(2.248,0.447){3}{\rule{1.567pt}{0.108pt}}
\multiput(573.00,764.17)(7.748,3.000){2}{\rule{0.783pt}{0.400pt}}
\put(584,768.17){\rule{2.500pt}{0.400pt}}
\multiput(584.00,767.17)(6.811,2.000){2}{\rule{1.250pt}{0.400pt}}
\put(596,770.17){\rule{2.300pt}{0.400pt}}
\multiput(596.00,769.17)(6.226,2.000){2}{\rule{1.150pt}{0.400pt}}
\put(607,771.67){\rule{2.650pt}{0.400pt}}
\multiput(607.00,771.17)(5.500,1.000){2}{\rule{1.325pt}{0.400pt}}
\put(618,772.67){\rule{2.891pt}{0.400pt}}
\multiput(618.00,772.17)(6.000,1.000){2}{\rule{1.445pt}{0.400pt}}
\put(630,773.67){\rule{2.650pt}{0.400pt}}
\multiput(630.00,773.17)(5.500,1.000){2}{\rule{1.325pt}{0.400pt}}
\put(641,774.67){\rule{2.650pt}{0.400pt}}
\multiput(641.00,774.17)(5.500,1.000){2}{\rule{1.325pt}{0.400pt}}
\put(675,774.67){\rule{2.650pt}{0.400pt}}
\multiput(675.00,775.17)(5.500,-1.000){2}{\rule{1.325pt}{0.400pt}}
\put(686,773.67){\rule{2.891pt}{0.400pt}}
\multiput(686.00,774.17)(6.000,-1.000){2}{\rule{1.445pt}{0.400pt}}
\put(698,772.17){\rule{2.300pt}{0.400pt}}
\multiput(698.00,773.17)(6.226,-2.000){2}{\rule{1.150pt}{0.400pt}}
\put(709,770.67){\rule{2.650pt}{0.400pt}}
\multiput(709.00,771.17)(5.500,-1.000){2}{\rule{1.325pt}{0.400pt}}
\multiput(720.00,769.95)(2.472,-0.447){3}{\rule{1.700pt}{0.108pt}}
\multiput(720.00,770.17)(8.472,-3.000){2}{\rule{0.850pt}{0.400pt}}
\put(732,766.17){\rule{2.300pt}{0.400pt}}
\multiput(732.00,767.17)(6.226,-2.000){2}{\rule{1.150pt}{0.400pt}}
\multiput(743.00,764.94)(1.505,-0.468){5}{\rule{1.200pt}{0.113pt}}
\multiput(743.00,765.17)(8.509,-4.000){2}{\rule{0.600pt}{0.400pt}}
\multiput(754.00,760.95)(2.472,-0.447){3}{\rule{1.700pt}{0.108pt}}
\multiput(754.00,761.17)(8.472,-3.000){2}{\rule{0.850pt}{0.400pt}}
\multiput(766.00,757.94)(1.505,-0.468){5}{\rule{1.200pt}{0.113pt}}
\multiput(766.00,758.17)(8.509,-4.000){2}{\rule{0.600pt}{0.400pt}}
\multiput(777.00,753.94)(1.505,-0.468){5}{\rule{1.200pt}{0.113pt}}
\multiput(777.00,754.17)(8.509,-4.000){2}{\rule{0.600pt}{0.400pt}}
\multiput(788.00,749.93)(1.267,-0.477){7}{\rule{1.060pt}{0.115pt}}
\multiput(788.00,750.17)(9.800,-5.000){2}{\rule{0.530pt}{0.400pt}}
\multiput(800.00,744.93)(0.943,-0.482){9}{\rule{0.833pt}{0.116pt}}
\multiput(800.00,745.17)(9.270,-6.000){2}{\rule{0.417pt}{0.400pt}}
\multiput(811.00,738.93)(1.155,-0.477){7}{\rule{0.980pt}{0.115pt}}
\multiput(811.00,739.17)(8.966,-5.000){2}{\rule{0.490pt}{0.400pt}}
\multiput(822.00,733.93)(1.033,-0.482){9}{\rule{0.900pt}{0.116pt}}
\multiput(822.00,734.17)(10.132,-6.000){2}{\rule{0.450pt}{0.400pt}}
\multiput(834.00,727.93)(0.798,-0.485){11}{\rule{0.729pt}{0.117pt}}
\multiput(834.00,728.17)(9.488,-7.000){2}{\rule{0.364pt}{0.400pt}}
\multiput(845.00,720.93)(0.798,-0.485){11}{\rule{0.729pt}{0.117pt}}
\multiput(845.00,721.17)(9.488,-7.000){2}{\rule{0.364pt}{0.400pt}}
\multiput(856.00,713.93)(0.758,-0.488){13}{\rule{0.700pt}{0.117pt}}
\multiput(856.00,714.17)(10.547,-8.000){2}{\rule{0.350pt}{0.400pt}}
\multiput(868.00,705.93)(0.692,-0.488){13}{\rule{0.650pt}{0.117pt}}
\multiput(868.00,706.17)(9.651,-8.000){2}{\rule{0.325pt}{0.400pt}}
\multiput(879.00,697.93)(0.692,-0.488){13}{\rule{0.650pt}{0.117pt}}
\multiput(879.00,698.17)(9.651,-8.000){2}{\rule{0.325pt}{0.400pt}}
\multiput(890.00,689.93)(0.669,-0.489){15}{\rule{0.633pt}{0.118pt}}
\multiput(890.00,690.17)(10.685,-9.000){2}{\rule{0.317pt}{0.400pt}}
\multiput(902.00,680.93)(0.611,-0.489){15}{\rule{0.589pt}{0.118pt}}
\multiput(902.00,681.17)(9.778,-9.000){2}{\rule{0.294pt}{0.400pt}}
\multiput(913.00,671.92)(0.547,-0.491){17}{\rule{0.540pt}{0.118pt}}
\multiput(913.00,672.17)(9.879,-10.000){2}{\rule{0.270pt}{0.400pt}}
\multiput(924.00,661.92)(0.600,-0.491){17}{\rule{0.580pt}{0.118pt}}
\multiput(924.00,662.17)(10.796,-10.000){2}{\rule{0.290pt}{0.400pt}}
\multiput(936.00,651.92)(0.496,-0.492){19}{\rule{0.500pt}{0.118pt}}
\multiput(936.00,652.17)(9.962,-11.000){2}{\rule{0.250pt}{0.400pt}}
\multiput(947.00,640.92)(0.496,-0.492){19}{\rule{0.500pt}{0.118pt}}
\multiput(947.00,641.17)(9.962,-11.000){2}{\rule{0.250pt}{0.400pt}}
\multiput(958.00,629.92)(0.496,-0.492){21}{\rule{0.500pt}{0.119pt}}
\multiput(958.00,630.17)(10.962,-12.000){2}{\rule{0.250pt}{0.400pt}}
\multiput(970.58,616.77)(0.492,-0.543){19}{\rule{0.118pt}{0.536pt}}
\multiput(969.17,617.89)(11.000,-10.887){2}{\rule{0.400pt}{0.268pt}}
\multiput(981.58,604.62)(0.492,-0.590){19}{\rule{0.118pt}{0.573pt}}
\multiput(980.17,605.81)(11.000,-11.811){2}{\rule{0.400pt}{0.286pt}}
\multiput(992.58,591.79)(0.492,-0.539){21}{\rule{0.119pt}{0.533pt}}
\multiput(991.17,592.89)(12.000,-11.893){2}{\rule{0.400pt}{0.267pt}}
\multiput(1004.58,578.47)(0.492,-0.637){19}{\rule{0.118pt}{0.609pt}}
\multiput(1003.17,579.74)(11.000,-12.736){2}{\rule{0.400pt}{0.305pt}}
\multiput(1015.58,564.65)(0.492,-0.582){21}{\rule{0.119pt}{0.567pt}}
\multiput(1014.17,565.82)(12.000,-12.824){2}{\rule{0.400pt}{0.283pt}}
\multiput(1027.58,550.47)(0.492,-0.637){19}{\rule{0.118pt}{0.609pt}}
\multiput(1026.17,551.74)(11.000,-12.736){2}{\rule{0.400pt}{0.305pt}}
\multiput(1038.58,536.32)(0.492,-0.684){19}{\rule{0.118pt}{0.645pt}}
\multiput(1037.17,537.66)(11.000,-13.660){2}{\rule{0.400pt}{0.323pt}}
\multiput(1049.58,521.51)(0.492,-0.625){21}{\rule{0.119pt}{0.600pt}}
\multiput(1048.17,522.75)(12.000,-13.755){2}{\rule{0.400pt}{0.300pt}}
\multiput(1061.58,506.17)(0.492,-0.732){19}{\rule{0.118pt}{0.682pt}}
\multiput(1060.17,507.58)(11.000,-14.585){2}{\rule{0.400pt}{0.341pt}}
\multiput(1072.58,490.17)(0.492,-0.732){19}{\rule{0.118pt}{0.682pt}}
\multiput(1071.17,491.58)(11.000,-14.585){2}{\rule{0.400pt}{0.341pt}}
\multiput(1083.58,474.23)(0.492,-0.712){21}{\rule{0.119pt}{0.667pt}}
\multiput(1082.17,475.62)(12.000,-15.616){2}{\rule{0.400pt}{0.333pt}}
\multiput(1095.58,457.02)(0.492,-0.779){19}{\rule{0.118pt}{0.718pt}}
\multiput(1094.17,458.51)(11.000,-15.509){2}{\rule{0.400pt}{0.359pt}}
\multiput(1106.58,439.87)(0.492,-0.826){19}{\rule{0.118pt}{0.755pt}}
\multiput(1105.17,441.43)(11.000,-16.434){2}{\rule{0.400pt}{0.377pt}}
\multiput(1117.58,422.09)(0.492,-0.755){21}{\rule{0.119pt}{0.700pt}}
\multiput(1116.17,423.55)(12.000,-16.547){2}{\rule{0.400pt}{0.350pt}}
\multiput(1129.58,403.87)(0.492,-0.826){19}{\rule{0.118pt}{0.755pt}}
\multiput(1128.17,405.43)(11.000,-16.434){2}{\rule{0.400pt}{0.377pt}}
\multiput(1140.58,385.72)(0.492,-0.873){19}{\rule{0.118pt}{0.791pt}}
\multiput(1139.17,387.36)(11.000,-17.358){2}{\rule{0.400pt}{0.395pt}}
\multiput(1151.58,366.96)(0.492,-0.798){21}{\rule{0.119pt}{0.733pt}}
\multiput(1150.17,368.48)(12.000,-17.478){2}{\rule{0.400pt}{0.367pt}}
\multiput(1163.58,347.57)(0.492,-0.920){19}{\rule{0.118pt}{0.827pt}}
\multiput(1162.17,349.28)(11.000,-18.283){2}{\rule{0.400pt}{0.414pt}}
\multiput(1174.58,327.57)(0.492,-0.920){19}{\rule{0.118pt}{0.827pt}}
\multiput(1173.17,329.28)(11.000,-18.283){2}{\rule{0.400pt}{0.414pt}}
\multiput(1185.58,307.68)(0.492,-0.884){21}{\rule{0.119pt}{0.800pt}}
\multiput(1184.17,309.34)(12.000,-19.340){2}{\rule{0.400pt}{0.400pt}}
\multiput(1197.58,286.57)(0.492,-0.920){19}{\rule{0.118pt}{0.827pt}}
\multiput(1196.17,288.28)(11.000,-18.283){2}{\rule{0.400pt}{0.414pt}}
\multiput(1208.58,266.41)(0.492,-0.967){19}{\rule{0.118pt}{0.864pt}}
\multiput(1207.17,268.21)(11.000,-19.207){2}{\rule{0.400pt}{0.432pt}}
\multiput(1219.58,245.54)(0.492,-0.927){21}{\rule{0.119pt}{0.833pt}}
\multiput(1218.17,247.27)(12.000,-20.270){2}{\rule{0.400pt}{0.417pt}}
\multiput(1231.58,223.26)(0.492,-1.015){19}{\rule{0.118pt}{0.900pt}}
\multiput(1230.17,225.13)(11.000,-20.132){2}{\rule{0.400pt}{0.450pt}}
\multiput(1242.58,201.26)(0.492,-1.015){19}{\rule{0.118pt}{0.900pt}}
\multiput(1241.17,203.13)(11.000,-20.132){2}{\rule{0.400pt}{0.450pt}}
\multiput(1253.58,179.54)(0.492,-0.927){21}{\rule{0.119pt}{0.833pt}}
\multiput(1252.17,181.27)(12.000,-20.270){2}{\rule{0.400pt}{0.417pt}}
\multiput(1265.58,157.11)(0.492,-1.062){19}{\rule{0.118pt}{0.936pt}}
\multiput(1264.17,159.06)(11.000,-21.057){2}{\rule{0.400pt}{0.468pt}}
\multiput(1276.58,134.11)(0.492,-1.062){19}{\rule{0.118pt}{0.936pt}}
\multiput(1275.17,136.06)(11.000,-21.057){2}{\rule{0.400pt}{0.468pt}}
\multiput(1287.58,111.40)(0.492,-0.970){21}{\rule{0.119pt}{0.867pt}}
\multiput(1286.17,113.20)(12.000,-21.201){2}{\rule{0.400pt}{0.433pt}}
\multiput(1299.58,87.96)(0.492,-1.109){19}{\rule{0.118pt}{0.973pt}}
\multiput(1298.17,89.98)(11.000,-21.981){2}{\rule{0.400pt}{0.486pt}}
\put(652.0,776.0){\rule[-0.200pt]{5.541pt}{0.400pt}}
\end{picture}
\caption{{
 The free energy excess $\Delta F/\delta p_d^2$
as a function of
$q_{min}^2=\l({3\ov 4}h_2^2\r)^{1/3}$ in two cases. Upper curve (a);
$q_{min}^1=0.027$, $h_1=0.0038$, $\Delta F=0$ for $q_{min}^2=0.027$.
Lower curve (b):  $q_{min}^1=h_1=0$ $\Delta F=0$ for $q_{min}^2=0$.
 The symmetry
of $\Delta F$ as a function of $q_1$ and $q_2$ implies that two curves are
equal in the opposite extremes.
}}
\end{figure}
 We also solved analytically the equations in the two limit cases
(1) $h_2=0$ and (2)
$h_2=h_1+\delta h$ with $\delta h << h_1$. In this
last case we just computed $\D F$ to the first order in $\delta h$.
The results for these two cases are:
\be
\D F=
\l\{
\begin{array}{ll}
\l({2187\ov 32}\r)^{1/3}
\; \del p_d^2\; h_1^{8/3}/q_{max} & h_2=0\\
{1\ov\sqrt{2}}\del p_d^2\;h_1 \;\del h & \del h<<h_1
\end{array}
\r.
\label{bottom}
\ee

In all cases the variational parameters turned out to be consistent
with the hypothesis of being of order $\delta p_d$ and with the positivity
condition (\ref{semi}), e.g. in the case $h_2=0$, $h_1\ne 0$, $x_m=0$,
we found $\del q_1^2=-\del p_d q_{min}/(
4 q_{max} - 3 q_{min})$ and $ \del p^1=\del p^2=\del p_d$ the other variables
being zero.

The computation of case (2) for  small
field difference $\delta h=h_2-h_1$ follows a very similar scheme.
 In this case we perturb around the
solution of the problem with $\del h=0$  ({\ref{2.8},\ref{2.9}})
with $q_{min} \leq p_d \leq q_{max}$.
Without entering in the details of the solution,
which is similar to the one of the
previous case, we just give the result.
Under the (self-consistent) hypothesis that all the variations are of
order $\del h$ one finds that the free energy excess is of order  $\del h^2$.
We get
\be
\Delta F={2\ov 3} \del h^2 (p_d-q_{min}) (p_d+q_{min}){( 3 p_d^2 +3 p_d q_{min}
+2
q_{min}^2)\ov p_d^2+p_d q_{min}+2 q_{min}^2  }
\label{middle}
\ee
note that, as it should be, the free energy  excess is zero for $p_d=q_{min}$.

Very similar paths can be followed  to study the  case (3) of
chaos with temperature.
We found  here for the free-energy excess
\be
\Delta F=\del p_d^2 {(T_1-T_2)^4\ov \tau_1}.
\label{top}
\ee
It is worth noticing that eq. (\ref{top}) is derived from the truncated model
in a magnetic field $h_1=h_2=h$ but does not depend on $h$.
As it was noticed by Kondor and V\'egs\"o \cite{Kondor2} the truncated model
presents a spurious instability in the fluctuation matrix. The result
(\ref{top}) shows that our calculation is insensitive to this instability.

Formulae (\ref{bottom},\ref{middle},\ref{top}) can be used to estimate the
probability distribution for an overlap $p_d$ among states at different
parameters in finite systems via the relation
${\cal P} (p_d) \sim  exp\; (-N\; \D F)$. This relation
allow for tests of  (\ref{bottom},\ref{middle},\ref{top}) in
numerical simulations.

{\bf Finite dimensions}

Let us now briefly comment about the relevance of our results for
finite dimensional spin glasses.

In addition to the spin-spin correlations an important quantity in
finite dimension is
 the correlation overlap function
$\overline {<S_iS_j>_1<S_iS_j>_2}$.
In a chaotic situation this  decays exponentially
with  a characteristic length $\xi_{1,2}$ for large  $|i-j|$.
This quantity was studied in
\cite{Kondor1,Kondor2} where it was found that, when $d>8$
\beqna\label {corrle}
\xi _{0,h} \;\sim \; h^{-2/3} &{\rm and}&
\xi _{T_1,T_2} \;\sim \; |T_1-T_2|^{-1}.
\eeqna
 This behaviour was confirmed in
 numerical simulations  by Ritort \cite{rit}.
The sensitivity
to small variations of an external parameter $X$ is characterized
by a "chaos exponent" $\zeta$,
in $\xi_{X_1,X_2} \sim |X_1-X_2|^{-1/ \zeta}$, first considered in
the framework
 of the scaling theory of Bray and Moore \cite{bray}
 and the droplet theory of Fisher and Huse \cite{huse}.

These results (\ref {corrle}) may be compared with ours
via the relation found in \cite{rit}
\beqna\label {pd}
N[\delta p_d^2]_{av} \; \sim \; \xi ^4
\eeqna
where $[\cdots]_{av}$ denotes the average with respect to
the distribution function of $p_d$,
${\cal P} (p_d) \sim  exp\; (-N\; \D F)$.
Upon substituting our results for the free energy excess
$\D F$ one recovers the
results  (\ref{corrle}) for the
dimension independent exponent $\zeta$ in the corresponding
cases, in the  case of
two non-zero magnetic fields, for small $|h_1-h_2|=\del h$
our result is:
\beqna
\xi_{h_1,h_1+\delta h} &\sim & (h_1\; \delta h)^{-1/4}.
\eeqna

In lower dimensions it is possible to determine $\xi _{T_1,T_2}$
\cite{bray,huse,ney} and $\D F$ \cite{hilh} within the scaling theories.
The differences with mean-field are that the
relation between the two is different from
(\ref{pd}) that gives $\D F \sim \del p_d^2 |T_1-T_2|^{4/\zeta}$,
 the exponent $\zeta$ now depends on dimension
and there are two regimes in temperature with
two different behaviours of the two quantities mentioned \cite{ney}
(one is the low-temperature phase, the other is the critical region).
It could be interesting to see whether the latter happens in
mean-field too.

\section{Conclusions}
We have studied in this paper the correlations among states
 at different magnetic fields and temperatures in some
 spin glass models. In the
REM and in the GREM it is  absent chaos with temperature,
 while there is chaos with magnetic field.
 This is understood  in simple terms
based on the ultrametric construction of temperature independent
trees. As soon as a temperature dependence is assumed, considering
correlated but not identical energy levels for different temperatures,
chaos is present. This could provide a possible mechanism for chaos
production in microscopic models.

In the SK model near $T_c$ we find that a free energy excess has to be paid to
constrain two systems to have an overlap greater than the one corresponding to
zero correlations, if the magnetic fields or the temperatures in the two
systems are different. This implies that all the possible
couples of  states with different external parameters and
 free energy density
equal to the one of the states dominating the partition function, have
minimal correlations.
The scenario we find has implication on the physical
picture of the low
temperature phase of the model.
The hypothesis  of successive bifurcations of the ultrametric tree as
the temperature is lowered
\cite{dotsen}
is incompatible with our results.

Let us conclude commenting on the fact that
temperature cycling experiments in spin glass off-equilibrium
relaxation \cite{dyn} show  strong correlations
in the  dynamics at different temperatures on finite time scales.
If the physics of experimental spin glasses were similar to that of the SK
model in this respect, one could expect these correlations eventually to decay
to zero for large times.
It would be very interesting in this context to test
the finite time  behavior  of the SK model in simulated
temperature cycling experiments.

\section*{Acknowledgements}
We thank  H.J. Hilhorst, I. Kondor, J. Kurchan, M. M\'ezard,
 G. Parisi, F. Ritort, M.A. Virasoro
for interesting  discussions.

\section*{Appendix}
In this appendix we show that no ultrametric solution of the
kind discussed in section 2 exists for the SK model near $T_c$ (\ref{F}).
We will show this by absurd, assuming an ultrametric solution
with $q_s(x)$ and $p(x)$ continuous in $x$. The discussion is done in
the case of the complete model,
 the same argument could also be applied to the
truncated model, with the same conclusions.

We discuss the case $h_1=h_2=0$ and different temperatures,
a similar (and simpler) prove leads to the conclusion that there
is chaos with magnetic field.
Let us write the variational equations for the complete model
 considering generic values of $w,u,y,v$:
\be
2\tau_{rs} \bQ_{\a\b} +
w (\bQ^2)_{\a\b} +
{2\ov 3} u \bQ_{\a\b}^3 -
y\sum_\gamma[\bQ_{\a\gamma}^2+\bQ_{\beta\gamma}^2] \bQ_{\a\b} +
v(\bQ^3)_{\a\b}=0.
\label{AA}
\ee
Plugging in the Parisi form for the matrices $Q_s$ and $P$ one get a set of
coupled integral equations
for the functions $q_s(x)$ and $p(x)$ that can
be solved by repeated differentiation with respect to $x$.

For future reference we write the solution of the free case
\cite{dedoko} at a temperature
$\tau=(1-T^2)/2$.
\be
q_F(x)=
\l\{
\begin{array}{ll}
{w\ov 2u}{x\ov\sqrt{1+{3v\ov 2u}x^2}} & x< \ol{x}
\\
q(1) & x\geq \ol{x}
\end{array}
\r.
\ee
Where $q(x)$ is continuous in $\ol{x}$ and $q(1)$ is specified by the equation
\be
2\tau+2y\la q^2\ra -2w q(1)+(3v +2u)q(1)^2=0.
\ee

In order to solve the problem we have to compute the Parisi
functions associated to the various terms
of (\ref{AA}). In particular we need to compute the functions associated to
\be\bQ^2=
\l(
\begin{array}{cc}
Q_1^2+P^2 & P(Q_1+Q_2)
\\
P(Q_1+Q_2) & Q_2^2+P^2
\end{array}
\r)
\ee
and
\be
\bQ^3=
\l(
\begin{array}{cc}
Q_1^3+P^2(2 Q_1+Q_2) & P^3 +P(Q_1^2+Q_2^2+Q_1 Q_2)
\\
 P^3 +P(Q_1^2+Q_2^2+Q_1 Q_2)
 & Q_2^3+P^2( Q_1+2Q_2)
\end{array}
\r).
\ee
To do that let us remind that the
eigenvalues associated to a Parisi matrix $A\to (a_d,a(x))$ are:
\be
\ll_0=a_d-\la a\ra \;\;\;{\rm with \;\;\; multiplicity} \;\;\; 1
\ee
\be
\ll(x)=a_d-x a(x)-\int_x^1
 \d y \; a(y) \;\;\;{\rm with \;\;\; multiplicity} \;\;\;
-n{\d x\ov x^2}.
\label{rel}
\ee
Observing that
\be
\ll'(x)=-xa'(x)
\label{derr}
\ee
one can invert the relation (\ref{rel})
and get
\be
a(x)=a(1)+\int_x^1 \d y \; {\ll'(y)\ov y}
\label{inv}
\ee
Let us denote $\ll_1(x),\ll_2(x),\ll_p(x)$ the eigenvalues associated with
$Q_1,Q_2,P$ respectively. The eigenvalues associated to
$\bQ^2$ will be:
\beqna
&Q_s^2+P^2\to \ll_s^2(x)+\ll_p^2(x)= {^2\L_s(x)}&
\\
&P(Q_1+Q_2)\to \ll_p(x)[\ll_1(x)+\ll_2(x)]= {^2\L_p(x)}
\eeqna
($s=1,2$) the corresponding functions can be obtained from (\ref{inv})
noting that as the magnetic field is zero $q_s(0)=p(0)=0$,
\beqna
^2A_s(x)=2\int_0^x \d y \; [ \ll_s(y) q_s'(y)+\ll_p(y)p'(y) ]
\\
^2A_p(x)=2\int_0^x \d y \;\{ p'(y)[ \ll_1(y)+\ll_2(y) ]
+[q'_1(y)+q'_2(y)]\ll_p(y)
\}
\eeqna
having made use of (\ref{derr}).
The derivative with respect to $x$ of these functions are:
\beqna
^2A_s'(x)=2[ \ll_s(x) q_s'(x)+\ll_p(x)p'(x) ]
\\
^2A_p'(x)=2\{ p'(x)[ \ll_1(x)+\ll_2(x) ] +[q'_1(x)+q'_2(x)]\ll_p(x)\}
\eeqna
In a completely analogous way one finds the functions
$^3A_s(x)$ and $^3A_p(x)$ and their derivatives:
\beqna
^3A_1'(x)=3 q_1' \ll_1^2+2p'\ll_p(2\ll_1+\ll_2)+\ll_p^2(2q_1'+q_2')
\label{3a1}
\\
^3A_p'(x)=3p'\ll_p^2+p'(\ll_1^2+\ll_2^2+\ll_1\ll_2)+\ll_p(2\ll_1q_1'
+2\ll_2q_2'+\ll_1q_2'+\ll_2 q_1')
\eeqna
The formula for $ ^3A_2'(x)$ is obtained from (\ref{3a1})
 interchanging the indices `1' and `2'.
Observing that
\be
\sum_\gamma [\bQ_{\a\gamma}^2+\bQ_{\b\gamma}^2]=
\l\{
\begin{array}{ll}
2[p_d^2 -<p^2> - <q_s^2> ] & r=s
\\
2[p_d^2 -<p^2>]-<q_1^2> - <q_2^2> & r\neq s
\end{array}
\r.
\ee
and that the functions associated to $\bQ_{\a\b}^3$ are just $q_s^3(x)$ and
$p^3(x)$ one finds that {\it the derivatives with respect to $x$}
of the saddle point equations read
\beqna\label{deriv}
[2\tau_1&-&2y (p_d^2 -<p^2> -<q_1^2>)]q_1'
 + 2 w [p'\ll_p+q_1' \ll_1]+
\nn
\\
& & v[3q_1'\ll_1^2+2p'\ll_p(2\ll_1+\ll_2)+\ll_p^2(2q_1'+q_2')]+2uq_1^2 q_1'=0
\\
& &
\nn
\eeqna
a similar equation with $1\leftrightarrow 2$, and
\beqna
& &[2
\tau_{12}
-y (2(p_d^2 -<p^2>) -<q_1^2>-<q_2^2>)]p'
\nn\\
& &+
w[p'(\ll_1+\ll_2))+(q_1'+q_2')\ll_p]+
\nn
\\
& &v[3\ll_p^2p'+p'(\ll_1^2+\ll_2^2+\ll_1\ll_2)+\ll_p(2\ll_1q_1'
+2\ll_2q_2'
\nn
\\
& &+\ll_1q_2'+\ll_2 q_1')]+2up^2p'=0.
\eeqna
Let us now study the possibility of an ultrametric solution.
Consider the `small $x$ region' defined in (\ref{um}), where
$q_1(x)=q_2(x)=p(x)=f(x)$, and suppose $f'(x)\neq 0$. One can the differentiate
repeatedly (\ref{deriv}) and find that
\be
f(x)=q_F(2x).
\ee
In the `large $x$ region', where $p(x)=p_d$ observing that
$\ll_p(x)=0$ there, one finds that the second equation
is automatically satisfied, while the first two equations reduce to
\be
q_s'[2\tau_s-2y(p_d^2 -<p^2> -<q_s^2>)+2w\ll_s+3v\ll_s^2+2uq_s^2]=0.
\label{grande}
\ee
that is, we get two un-coupled equations for $q_1$ and $q_2$ in this region.
Again by repeated differentiation, we find that if $q_s'\neq 0$ then
$q_s(x)=q_F(x)$.
Using the assumption of continuity we find
\be
q_s(x)=
\l\{
\begin{array}{ll}
q_F(2x) & x\leq x_0/2
\\
p_d & x_0/2 < x\leq x_0
\\
q_F(x)& x_0 < x \leq \ol{x}_s
\\
q_s(1) & \ol{x}_s<x\leq 1
\end{array}
\r.
p(x)=
\l\{
\begin{array}{ll}
q_F(2x) & x\leq x_0/2
\\
p_d & x_0/2 < x\leq 1
\end{array}
\r.
\ee
The only free parameter at this level are the values $q_1(1)$ and $q_2(1)$.
These can be fixed e.g. considering eq. (\ref{grande}) in $x=\ol{x}_s$
which gives
\be
2\tau_s+2y
\l[ \int_0^{\ol{x}_s} \d x \; q_F(x)^2 +(1-\ol{x}_s)q_s(1)^2 \r]
-2w q_s(1)+(3v +2u)q_s(1)^2=0.
\ee
showing that $q_s(1)$ is equal to the value $q_F(1)$ corresponding to
$\tau=\tau_s$.
If now one inserts the resulting functions in equations (\ref{deriv})
one finds the absurd
\be
(p_d-<p>)(<q_1>-<q_2>)=0
\ee
showing the inconsistency of the  hypothesis of an ultrametric solution
except for the trivial one with  $p_d= 0$.

\end{document}